\documentstyle[12pt, bezier]{article}
\begin{document}
\input FEYNMAN
\renewcommand{\theequation}{\thesection.\arabic{equation}}

\vspace*{2cm}
\begin{center}
{\LARGE\bf Low x Physics}
\end{center}
\begin{center}

{\LARGE\bf and Perturbative Regge Asymptotics\footnote{ Talk at the
Tennessee International Symposium on Radiative Corrections, June 27 -
July 1, 1994 } }

\end{center}
\vspace{0.5cm}
\begin{center}
{\large\bf R. Kirschner}\\{\large\it DESY - Institut f\"ur
Hochenergiephysik Zeuthen }\\{\large\it Platanenallee 6,
D-15735 Zeuthen, Germany }
\end{center}
\vspace*{5.0cm}
\begin{center}
{\bf Abstract }
\end{center}

Deep-inelastic scattering at small x is a new field of
QCD phenomenology which HERA experiments started to explore. The data
show the feature expected by the perturbative Regge asymptotics. We
describe the parton picture of structure function evolution both for
increasing $Q^2$ and for decreasing $x$ and discuss the leading
logarithmic approximation of QCD on which this picture is based.
The steep rise of the parton density towards smaller $x$ gets saturated
by parton recombination. An essential improvement of the leading
logarithmic approximation has to be worked out in order to satisfy the
unitarity conditions and in this way to describe the parton
recombination in perturbative QCD. We review some ideas of the
perturbative Regge asymptotics and some recent theoretical results about
multi-reggeon exchange in QCD.

\newpage

\baselineskip=14pt
\section{Deep-inelastic scattering at small $x$}
The parton distribution obtained from deep-inelastic structure functions
enter the expression for all hard processes. At high energies the
small-$x$ region of the distributions gives the dominant contribution.
The measurements at HERA \cite{H1F2} \cite{ZF2} of $F_2(x,Q^2)$ at
values
of $x$ down to $10^{-4}$ reduce essentially the uncertainties in this
region. Moreover,  HERA provides the first experimental information
about an unexplored field of QCD phenomenology.

At small $x$ the virtual Compton amplitude, the forward imaginary part
of which determines the structure functions, is in the Regge regime.
Indeed, we have for the c.m. energy squared of the virtual photon and
the proton
\begin{equation}
s \simeq {Q^2 \over x} \gg Q^2 \gg m_p^2 .
\end{equation}
It is a peculiar Regge limit because of  $Q^2$ being much larger than
the hadronic scale (represented by the proton mass $m_p$). This
provides
the advantage that an essential part of the deep-inelastic interaction
can be calculated perturbatively also at small $x$.

The perturbative Regge asymptotics in non-abelian gauge theories has
been a topic of theoretical investigation already 20 years ago.
In the leading logarithmic approximation the asymptotics for vacuum
quantum number exchange is determined \cite{BFKL} \cite{ChL} by a branch
point singularity
at the angular momentum $ 1 + \omega_{0}, \ \ \omega_{0} = {g^2 N \over
2 \pi^2 } 2 \ln 2 $. The perturbative
contributions lead to a power-like behaviour of the structure function
at small $x$,
\begin{equation}
F_2 (x,Q^2) \sim \left ({1 \over x}\right )^{\omega_0 } .
\end{equation}
Identifying ${g^2 \over 4\pi }$ with $\alpha_S (10 GeV^2)$ and choosing
$N = 3$ one estimates
$\omega_0 = 0.5 $. This is a much steeper increase towards smaller $x$
compared to the expectations from hadron Regge phenomenology (pomeron
intercept at 1.08 ) or to the asymptotics induced by the usual
$Q^2$-evolution of structure functions ($\sim \exp (B ( - \ln x )^{1/2})
$, called $r$-evolution below). The data give evidence that this
increase,
Eq. (2), really takes place. The seemingly smaller power in the data at
lower $Q^2$ is a subject of a more detailed discussion.

 The perturbative Regge asymptotics is expected to show up not only in
the $x$-dependence but also in certain features of the hadronic final
state, which is a subject of experimental \cite {UO} \cite{ET} and
theoretical \cite{AM} \cite{BdR} \cite{Kw} investigations.

\section{The parton picture}
Let us draw the parton picture of the structure function evolution
before we discuss some details about how it emerges from QCD.

The parton hit by the virtual photon can be considered as emerging from
an evolution starting from the hadron constituents and proceeding by
parton radiation. At each step $\ell$ of the radiation process the
transverse momentum $\kappa_{\ell }$ and the longitudinal momentum
fraction $x_{\ell }$ of the parton change. $x$ is the value of $x_{\ell
}$ at the final step ($\ell = n ) $ of the evolution and $Q$ is the
upper bound for $\vert \kappa \vert $. If $x$ is small then there is
room for this process  to evolve both in $r = \ln ( \vert \kappa^2
\vert
/ m_p^2 ) $ as well as in $y = - \ln x $. The $y-r$ plot Fig. 1 serves
as an useful illustration \cite{EL}.
The features of the evolution in $r$ are well known. At higher steps in
this evolution the partons appear in higher multiplicitiy and at higher
resolution (a reasonable measure of which is just $r$). The new partons
cover a negligible fraction of the transverse space occupied by the
parent parton.

The evolution in $y$ proceeds without essential changes in the
resolution. The new
partons appear with about the same size as the parent parton. According
to Eq.(2) the multiplicity grows exponentially in $y$. The partons tend
to cover the space inside the proton. The rapid growth of the
multiplicity in the first stage of the $y$-evolution becomes saturated
by recombination effects before the partons come too close to each
other. The region where the recombination is essential is bounded by the
critical line in the $y-r$ plot Fig.1.

%Fig. 1

\unitlength=1.00mm
%\special{em:linewidth 0.4pt}
\linethickness{0.4pt}
\begin{picture}(148.00,100.00)
\put(20.00,15.00){\line(1,0){120.00}}
\put(5.00,15.00){\line(1,0){15.00}}
\put(55.00,30.00){\line(0,1){48.00}}
\put(55.00,80.00){\circle{3.00}}
\put(55.00,29.00){\circle{3.00}}
\put(2.00,55.00){\line(1,0){11.00}}
\put(22.00,56.00){\line(6,1){12.00}}
\put(45.00,61.00){\line(4,1){14.00}}
\put(70.00,68.00){\line(5,2){13.00}}
\put(93.00,77.00){\line(2,1){11.00}}
\put(11.00,90.00){$y$}
\put(140.00,6.00){$r$}
\put(20.00,13.00){\line(0,1){84.00}}
\put(19.00,6.00){$0$}
\bezier{848}(30.00,80.00)(55.00,-23.00)(80.00,80.00)
\end{picture}
\unitlength=0.0035mm

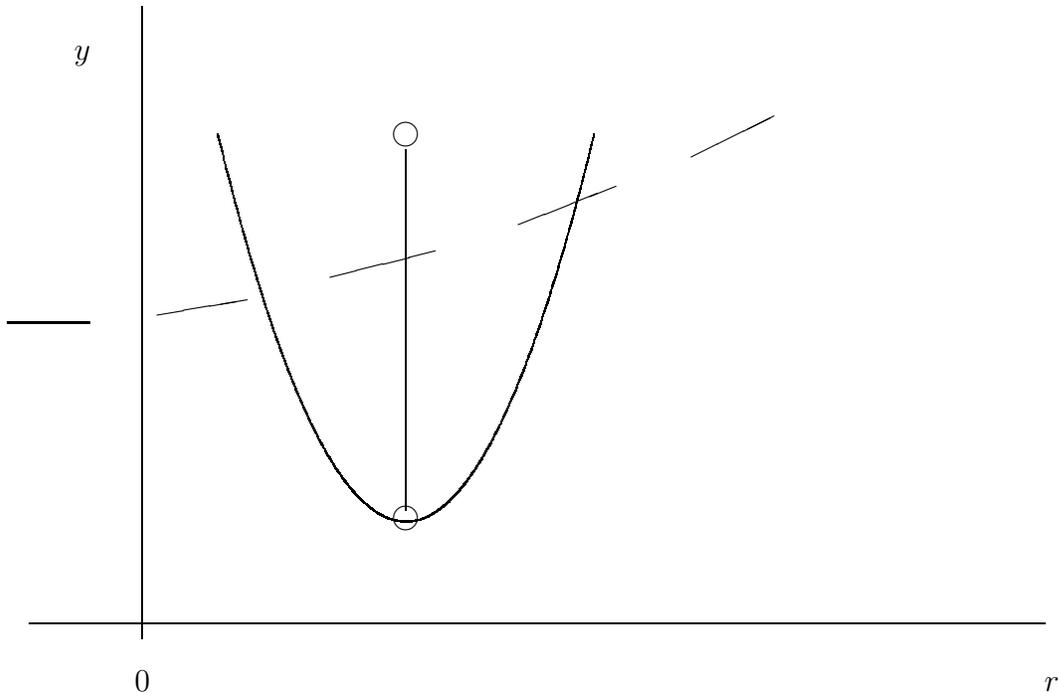
{\tenrm Fig.1: The evolution of the parton density illustrated in the
$y-r$
plot. Left to the $y$ axis is the non-perturbative region. The dashed
line indicates the critical line and the parabola shows the $r$-range
broadening due to diffusion in the $y$-evolution.}
\vspace{1cm}

During the evolution in $y$ the value of $r$ undergoes a random walk and
may run into the non-perturbative region ($r < 0 $). This means in
particular that the value of the parton density at given $r$ and $y$
becomes influenced by non-perturbative contributions  if $y $ becomes
large enough even if $r$ is far from the non-perturbative region.

In both the $r$-evolution and the $y$-evolution the process starts
with the proton constituents inside
the non-perturbative region. Measuring a jet with $\kappa_j, x_j $ in
the hadronic final state one detects an intermediate state of the
evolution in the perturbative region (provided $\kappa_j$ is not small).
In the case $x_j \gg x, \vert \kappa_j^2 \vert \sim Q^2 $ this
measurement selects a nearly ideal perturbative $y$-evolution, all
features of which are calculable \cite{AM}.

\section{The leading logarithmic approximation}
In certain kinematical regions large perturbative correction in each
order arise from the integration over $\kappa $ or $x$ of (real or
virtual) partons, if the range in $\kappa$ is large or if the range in
$x$ extends to small values and if the whole range contributes
uniformly. In this case the corresponding integrals are approximately of
logarithmic form.

In the leading $\ln \kappa^2 $ approximation one picks up the
logarithmic contributions from the transverse momentum integral of each
loop. Graphs with non-logarithmic loops are neglected. Applied to the
structure functions at $r >  y $   this is the appropriate
approximation
which can be improved systematically. The leading contribution can be
represented as a sum of ladder graphs ( in an axial gauge, with
propagator and vertex corrections included). There is a strong ordering
of the transverse momenta in the loops which leads in particular to
trivial transverse momentum integrals. Summing these graphs leads to the
known GLAP equation \cite{GLAP}. The iterative structure of the ladders
implies in particular the following form of the structure functions,
exhibiting their universality.
\begin{equation}
F_2 (x,Q^2 ) = {d \sigma^{\gamma* j } \over d \hat t } \vert_{\hat t =
0} \otimes
D_{ji} ({x_1 \over x_2}, Q^2, Q_0^2) \otimes
F_i^{(0)}(x_2,Q_0^2)
\end{equation}
$\otimes$ abbreviates a convolution integral in the longitudinal
momentum fraction. A summation over the parton types $i,j$ is
understood.
The first factor represents the forward cross section for the virtual
photon - parton scattering . Replacing it by other hard scattering cross
sections allows to relate the structure functions  to distributions in
those processes. The second factor represents the $r$-evolution, the
solution of the GLAP equation with the initial distribution $\delta_{ij}
\delta ({x_1 \over x_2} -1) $ at $Q_0^2 $.

Since the first two factors are calculated from QCD all experimental
information about structure functions and hard processes can be reduced
to the distributions $F_i^{(0)} (x, Q_0^2) $ of partons in the hadron at
$Q_0^2$.

In the leading $\ln x$ approximation the integrations over $x$ and
$\kappa$ interchange their roles compared to the above case. For
structure functions it is the appropriate approximation at $y > r$.
Again the leading contribution can be represented as a sum of ladder
graphs, Fig. 2.

Now the lines in the $t$-channel direction represent
reggeized gluons and the interaction is determined by effective
vertices. A strong ordering holds now in the longitudinal momenta. More
precisely, the loop momenta $k_{\ell} $ obey the conditions of
multi-Regge kinematics which can be written using the Sudakov
decomposition,
\begin{equation}
k = \sqrt{\frac{2}{s}} \ \left (k_- \  q^{\prime } +  k_+ \ p \right ) +
\kappa, \ \ \ \ q^{\prime } = q - x p,
\end{equation}
($q$ and $p$ are the 4-momenta  of the photon and the proton), in the
following form
\begin{eqnarray}
&k_{+ n} \ll ... \ll k_{+ 1 } , \ \ \ \ \ \  \ \ k_{- n} \gg ... \gg
k_{- 1}, \cr
&k_{+ \ell} k_{- \ell} \ll \vert \kappa_{\ell }^2 \vert, \ \ \ \ \ \ \ \
s_{\ell } \equiv k_{+ \ell -1} k_{- \ell +1} \gg \vert \kappa_{\ell }^2
\vert,                                           \cr
&\prod_{\ell =1}^n s_{\ell } = s \prod_{\ell =2}^n \vert \kappa_{\ell }
-\kappa_{\ell -1} \vert^2.
\end{eqnarray}
\vspace{1cm}

%Fig. 2
\begin{picture}(50000,21000)
%\drawline\scalar[\N\REG](5000, 0)[3]
%\drawline\scalar[\N\REG](5000, 16000)[3]
%\put(5000,11000){\oval(3000,10000)}
\drawline\photon[\SW\FLIPPEDFLAT](8000,6000)[9]
\drawline\photon[\NW\FLIPPEDFLAT](8000,16000)[9]
\drawline\fermion[\E\REG](8000,6000)[3000]
\drawline\fermion[\N\REG](\pbackx,\pbacky)[10000]
\drawline\fermion[\W\REG](\pbackx,\pbacky)[3000]
\drawline\fermion[\S\REG](\pbackx,\pbacky)[10000]
\put(9000,10500){$ \Phi^{\gamma } $}
\drawline\photon[\E\REG](11000,8500)[24]
\drawline\photon[\E\REG](11000,13500)[24]
\drawline\photon[\N\CURLY](15000,8700)[4]
\drawline\photon[\N\CURLY](19000,8700)[4]
\drawline\photon[\N\CURLY](27000,8700)[4]
\drawline\photon[\N\CURLY](31000,8700)[4]
\put(15000,8500){\circle*{500}}
\put(13000,7000){$k_n$}
\put(19000,8500){\circle*{500}}
\put(17000,7000){$k_{n-1}$}
\put(27000,8500){\circle*{500}}
\put(29000,7000){$k_2 $}
\put(31000,8500){\circle*{500}}
\put(33000,7000){$k_1$}
\put(15000,13500){\circle*{500}}
\put(19000,13500){\circle*{500}}
\put(27000,13500){\circle*{500}}
\put(31000,13500){\circle*{500}}
\put(21500,11000){. \ . \ . }
\drawline\fermion[\N\REG](37000, 0)[6000]
\drawline\fermion[\N\REG](37000, 16000)[6000]
\drawline\fermion[\N\REG](36750, 0)[6000]
\drawline\fermion[\N\REG](36750, 16000)[6000]
\drawline\fermion[\N\REG](37250, 0)[6000]
\drawline\fermion[\N\REG](37250, 16000)[6000]
\put(37000,11000){\oval(3250,10000)}
\put(36500,10500){$ \Phi^{(0)} $}
\put(2000,-1000){$ q $}
\put(36500, -1000){$ p $}
%\put(40500,-1000){\sl Fig. 2 }

\end{picture}

\vspace{1cm}

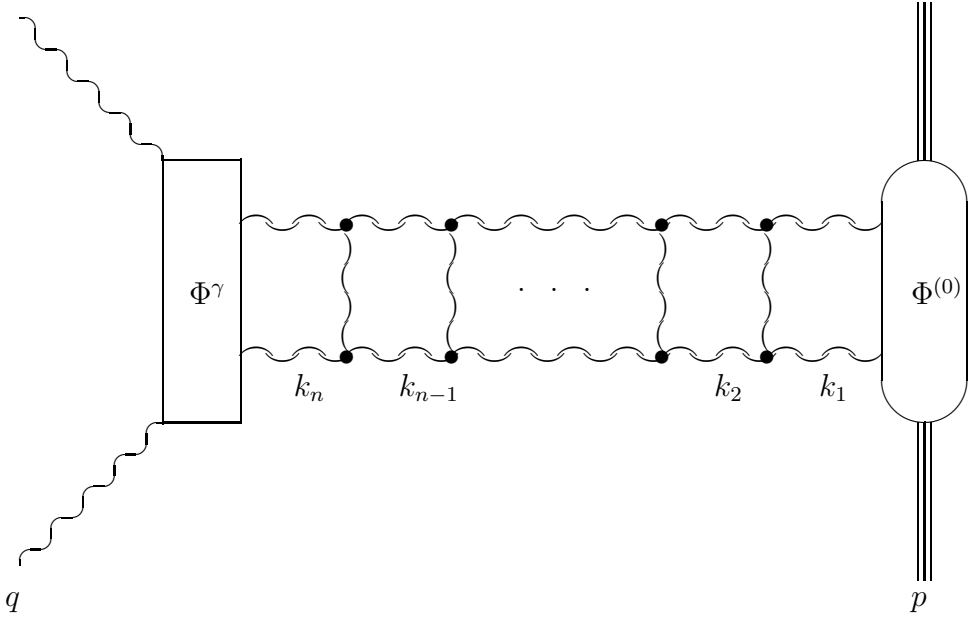
{\tenrm Fig. 2: \ \   Contribution to the structure function in the
leading $\ln x $ approximation. }
\vspace{1cm}

The longitudinal momentum integrals become trivial. The iterative
structure of the ladders leads to the following  form of the resulting
structure functions,
\begin{equation}
F_2 (x,Q^2 ) = \Phi^{\gamma*}({x \over x_1}, Q, \kappa ) \otimes
\tilde f ({x_1 \over x_2}, \kappa, \kappa_0 ) \otimes
\Phi^{(0)} (x_2, \kappa_0).
\end{equation}
Unlike in Eq. (3) here $\otimes $ denotes a transverse momentum
integration besides of a trivial $x$-integral. The universality of the
structure functions holds in the $y$-evolution with this modification,
the phenomenological relevance of which has been discussed recently
\cite{FR}. Here the first factor represents the impact factor coupling
(via a quark loop) the virtual photon to the gluon ladder. The second
factor describes the $y$-evolution. Because the first two factors are
calculated in QCD the non-trivial information is reduced to the third
factor, the hadron impact factor.

The $y$-evolution is calculated as the solution of the BFKL equation
\cite{BFKL} \cite{ChL}, a simple form of which (at vanishing momentum
transfer) is given by
\begin{equation}
{\partial \over \partial y} \tilde f (y, \kappa, \kappa_0) =
\frac{g^2 N}{(2 \pi)^3} \int {d^2 \kappa^{\prime } \over \vert \kappa -
\kappa^{\prime } \vert^2 }
{ 2 \vert \kappa^2 \vert \over \vert \kappa^{\prime 2} \vert }
\left ( \tilde f (y, \kappa^{\prime }, \kappa_0) -
\frac{1}{2}  \tilde f (y, \kappa, \kappa_0) \right ).
\end{equation}
The asymptotics of the solution at large $y$,
\begin{eqnarray}
&\tilde f (y, \kappa, \kappa_0) \sim
{ \vert \kappa \vert  \over (2 \pi D^2 y)^{1/2} }\ \
e^{\omega_0 y} \ \
\exp ({- \ln^2 (\vert \kappa^2 \vert / \vert \kappa_0^2 \vert ) \over
2 D^2 y }), \cr
&\omega_0 = {g^2 N \over 2 \pi^2 } 2 \ln 2,  \ \ \ \ \ \ \ \ \ \
D^2 = {g^2 N \over  \pi^2 } \ 7 \ \zeta (3),
\end{eqnarray}
exhibits the power-like increase in $\frac{1}{x} $ and the diffusion in
$r$ mentioned above.

Structure function parametrizations \cite{MRS} \cite{CTEQ} \cite{GRV}
rely on Eq. (3) and look for the optimal parton distributions
$F_i^{(0)}(x, Q_0^2)$ describing
all data (Fig. 3). The low-$x$ data are well described by assuming the
behaviour
Eq. (2). There is a parametrization approach \cite{GRV} which manages to
describe the
rise at small $x$ without assuming a rising initial parton distribution
like Eq. (2). The point is that the small $x$ asymptotics induced by the
GLAP equation $\sim \exp (B (- \ln x)^{1/2} ) $ can be close to the rise
originating from an input distribution like Eq. (2) at not too small $x$
by choosing the starting point $Q_{0}^2 $ correspondingly   (actually
$B^2 = {4 N \alpha_S
\over \pi } \ln {Q^2 \over Q_{0}^2 }$ ). However a choice of $Q_{0} $
deep
in the non-perturbative region which is necessary to describe the data
in that way cannot be justified.

In the small-$x$ region a complimentary approach to the parametrization
can be considered, relying on the $y$-evolution Eqs. (6), (7) and
parametrizing the hadron impact factor $ \Phi^{(0)}(x_0, \kappa )$ at
some initial $y_0 = \ln \frac{1}{x_0} $  as
a function of $\kappa$ (Fig. 3). The behaviour of this impact factor at
large
$\kappa^2$ has to be compatible with the GLAP equation. Some steps along
this line have been done \cite{Kw2}.

%Fig. 3

\unitlength=1mm
%\special{em:linewidth 0.4pt}
\linethickness{0.4pt}
\begin{picture}(147.00,100.00)
\put(20.00,23.00){\line(0,1){72.00}}
\put(3.00,25.00){\line(1,0){144.00}}
\put(15.00,88.00){$y$}
\put(136.00,20.00){$r$}
\put(20.00,20.00){$0$}
\put(5.00,37.00){\line(1,0){5.00}}

\put(5.00,37.00){\line(1,0){5.00}}

\put(15.00,37.00){\line(1,0){5.00}}

\put(25.00,37.00){\line(1,0){5.00}}

\put(35.00,37.00){\line(1,0){5.00}}

\put(45.00,37.00){\line(1,0){5.00}}

\put(55.00,37.00){\line(1,0){5.00}}

\put(65.00,37.00){\line(1,0){5.00}}

\put(75.00,37.00){\line(1,0){5.00}}

\put(85.00,37.00){\line(1,0){5.00}}

\put(95.00,37.00){\line(1,0){5.00}}

\put(105.00,37.00){\line(1,0){5.00}}

\put(115.00,37.00){\line(1,0){5.00}}

\put(125.00,37.00){\line(1,0){5.00}}

\put(5.00,37.40){\line(1,0){5.00}}

\put(5.00,37.40){\line(1,0){5.00}}

\put(15.00,37.40){\line(1,0){5.00}}

\put(25.00,37.40){\line(1,0){5.00}}

\put(35.00,37.40){\line(1,0){5.00}}

\put(45.00,37.40){\line(1,0){5.00}}

\put(55.00,37.40){\line(1,0){5.00}}

\put(65.00,37.40){\line(1,0){5.00}}

\put(75.00,37.40){\line(1,0){5.00}}

\put(85.00,37.40){\line(1,0){5.00}}

\put(95.00,37.40){\line(1,0){5.00}}

\put(105.00,37.40){\line(1,0){5.00}}

\put(115.00,37.40){\line(1,0){5.00}}

\put(125.00,37.40){\line(1,0){5.00}}

\put(15.00,40.00){$y_0$}
\put(40.00,23.00){\line(0,1){65.00}}
\put(40.40,25.00){\line(0,1){63.00}}
\put(40.00,20.00){$r_0$}
\end{picture}
\unitlength=0.0035mm

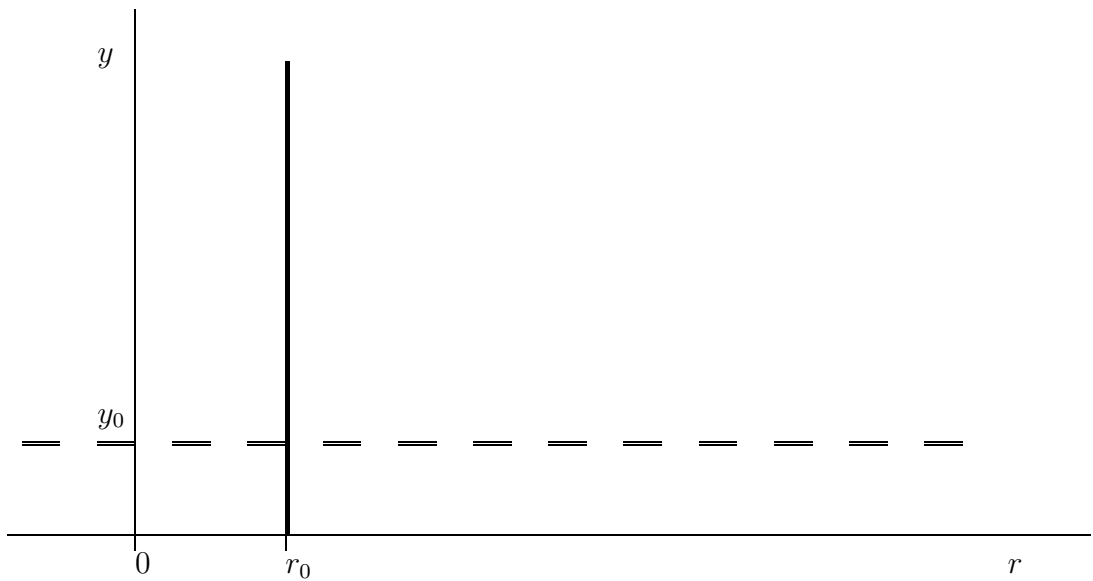
{\tenrm Fig. 3: The parametrization of parton densities illustrated in
the
$y-r$ plot. The initial values of the hadron structure function in the
$r$-evolution are given at the vertical (double) line through $r_0$. The
$y$-evolution starts from the hadron impact factor given on the
horizontal (dashed double) line through $y_0$. }

\vspace{2cm}

The special features of the $y$-evolution show up in several
characteristics of the hadronic final state. The inclusive jet
distributions have been mentioned above. The transverse energy flow in
the rapidity region away from the current jet has been observed
\cite{ET} to be more pronounced at small $x$. It turns out that these
data are not well described by particle production according to the
mechanism of the GLAP equation. A good description at small $x$ is
achieved when using a production mechanism based on the BFKL equation
\cite{Kw}.

The critical line, beyond which recombination is important, is estimated
by using the simple unitarity argument \cite{GLR} that the $\gamma^{*} p
$ cross section must be bounded by the geometric cross section. This
should still be true if $\gamma^{*} $ would interact strongly, i.e. if
$\alpha_{QED} $ is replaced by $\alpha_S(Q^2)$.
\begin{equation}
{\alpha_S(Q^2) \over \alpha_{QED} } \sigma^{\gamma^* p}  =
{\alpha_S(Q^2) \over Q^2} F_2 (x, Q^2) < \pi R_p^2.
\end{equation}
Substituting the result of the BFKL equation we obtain the limit of
applicability of this equation. Unitarity demands to go beyond the
leading logarithmic approximation. A minimal way to improve the
unitarity properties of this approximation will be discussed in the
last section.

An approach to unitarity improvent has been proposed by Gribov, Levin
and Ryskin \cite{GLR}. It amounts in including a quadratic term in the
evolution equation. Such a term accounts for the splitting of the gluon
ladder into two ladders, assuming a simple splitting vertex. The GLR
equation has been studied in detail \cite{GLReq}.

In QCD there are contributions with more than two reggeized guons in the
$t$-channel. In general they do not come in colour-singlet pairs. A
recent study \cite{BW} of the perturbative QCD asymptotics of the
$\gamma^* p $ high mass diffraction shows clearly that the transition
vertex of a two-reggeon state to a four-reggeon state in the $t$-channel
is more complicated than the one assumed in the GLR approach.

\section{The perturbative Regge asymptotics}

In the kinematics Eq. (1) the hard scattering is determined by a small
coupling. In the $y$-evolution ($ x_0 \rightarrow x \ll x_0 , Q_{0}^2
\rightarrow Q^2 \sim Q_{0}^2 $) the range of the transverse momentum
integral is localized around $Q_{0}$.  The diffusion in $r$ is a higher
order effect. Therefore one can start calculating with a fixed coupling
(at scale $Q_{0}^2 $ ). The effects of the renormalization group are to
be included in a next step \cite{KL}.

We shall discuss some basic ideas of the perturbative Regge asymptotics.
It is convenient to work with partial waves obtained by Mellin
transformation of the $x$-dependence. The Mellin variable $\omega $ is
related to the angular momentum $j$ by $j = 1+ \omega $.

 The exchanged gluons become reggeons. The trajectory is $1 + N \alpha_G
(\kappa ) $, where in the leading logarithmic approximation we have
\begin{equation}
\alpha_G (\kappa ) = {g^2 \over 2(2\pi)^3 }
\int { d^2\kappa^{\prime } \vert \kappa^2 \vert \over
\vert \kappa^{\prime 2} \vert \vert \kappa - \kappa^{\prime } \vert^2 }.
\end{equation}
The asymptotics of the vacuum exchange channel is dominated by the
exchange of an even number of reggeized gluons interacting by the
exchange of s-channel gluons, Fig. 4.

%Fig. 4
\begin{picture}(40000,21000)
\drawline\scalar[\E\REG](3000,4000)[16]
\drawline\scalar[\E\REG](3000,8000)[16]
\drawline\scalar[\E\REG](3000,12000)[16]
\drawline\scalar[\E\REG](3000,16000)[16]

\drawline\fermion[\N\REG](8000,4000)[4000]
\put(8000,4000){\circle*{500}}
\put(8000,8000){\circle*{500}}

\drawline\fermion[\N\REG](12000,4250)[3500]
\put(12000,4000){\circle*{500}}
\put(12000,12000){\circle*{500}}
\put(12000,8000){\oval(500,500)[r]}
\drawline\fermion[\N\REG](12000,8250)[3500]

\drawline\fermion[\N\REG](16000,4250)[3500]
\put(16000,4000){\circle*{500}}
\put(16000,16000){\circle*{500}}
\put(16000,8000){\oval(500,500)[r]}
\put(16000,12000){\oval(500,500)[r]}
\drawline\fermion[\N\REG](16000,8250)[3500]
\drawline\fermion[\N\REG](16000,12250)[3500]

\drawline\fermion[\N\REG](20000,8000)[4000]
\put(20000,8000){\circle*{500}}
\put(20000,12000){\circle*{500}}

\drawline\fermion[\N\REG](24000,8250)[3500]
\put(24000,8000){\circle*{500}}
\put(24000,16000){\circle*{500}}
\put(24000,12000){\oval(500,500)[r]}
\drawline\fermion[\N\REG](24000,12250)[3500]

\drawline\fermion[\N\REG](28000,4250)[3500]
\put(28000,4000){\circle*{500}}
\put(28000,12000){\circle*{500}}
\put(28000,8000){\oval(500,500)[r]}
\drawline\fermion[\N\REG](28000,8250)[3500]

\drawline\fermion[\N\REG](32000,12000)[4000]
\put(32000,12000){\circle*{500}}
\put(32000,16000){\circle*{500}}
\end{picture}

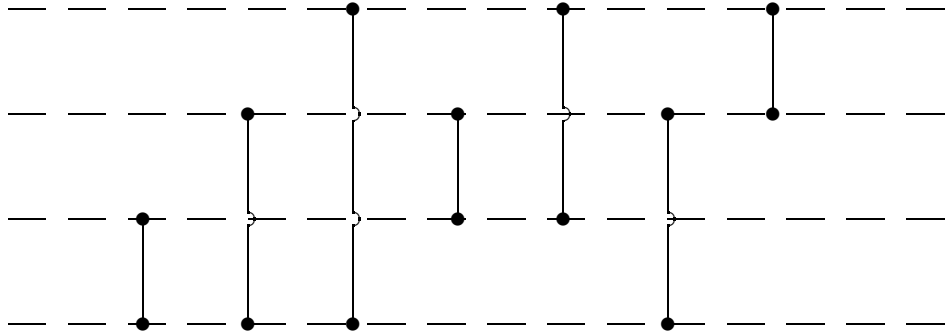
{\tenrm Fig. 4: \ \ Exchange of 4 reggeized gluons (dashed lines)
interacting via s-channel gluons (full lines). }
\vspace{2cm}

This interaction is determined
by effective vertices obtained by adding to the appropriate projection
of the
usual three-gluon vertex the contributions of brems\-strah\-lung.

%Fig. 5
\begin{picture}(30000,20000)
\drawline\fermion[\E\REG](3000,10000)[6000]
%\drawarrow[\W\ATBASE](4500,10000)
\drawline\photon[\N\REG](6000,10000)[3]
\put(6000,10000){\circle*{300}}
\put(4000,8000){$  \kappa $}

\put(8000,8000){$  \kappa^{\prime } $}
\put(6000,4000){\sl a}
\drawline\fermion[\E\REG](14000, 10000)[12000]
\drawline\photon[\N\REG](20000,10000)[3]
\put(20000,10000){\circle*{300}}
\put(20000,13000){\circle*{300}}

\drawline\fermion[\E\REG](14000, 13000)[12000]
%\drawarrow[\W\ATBASE](18000,10000)
\put(15000,8000){$    \kappa_1 $}
\put(15000,14000){$  \kappa_2 $}
\put(25000,8000){$  \kappa_1^{\prime } $}
\put(25000,14000){$  \kappa_2^{\prime } $}
\put(20000,4000){\sl b}

%\put(30000,0){\sl Fig. 5}
\end{picture}

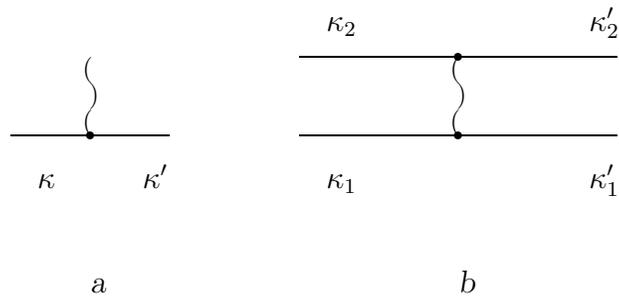
{\tenrm Fig. 5: \ \ a) Effective vertex of gluon emission from an
exchanged (reggeized) gluon. \newline
b) Interaction of two exchanged (reggeized) gluons.}

\vspace{2cm}

For
the two helicity states (represented by $\phi $ and $\phi^* $) of the
s-channel
gluon the leading logarithmic approximation of the effective vertex is
given by (compare Fig. 5a)
\begin{equation}
{\kappa^{\prime } \kappa^* \over (\kappa - \kappa^{\prime })} \phi^* +
{\kappa^{*} \kappa^{\prime } \over (\kappa^{*} - \kappa^{\prime *})}
\phi   .
\end{equation}
We represent the transverse momenta by complex numbers. Eq. (11)
implies
for the bare interaction kernel of two reggeized gluons (Fig. 5b)
\begin{equation}
{\cal H}^{(0)} (\kappa_1, \kappa_2; \kappa_1^{\prime }, \kappa_2^{\prime
}) =
\delta (\kappa_1 + \kappa_2 - \kappa_1^{\prime } - \kappa_2^{\prime })
{ \kappa_1 \kappa_1^{\prime *} \kappa_2^* \kappa_2^{\prime } + {\rm
c.c.} \over
\vert \kappa_1 - \kappa_1^{\prime } \vert^2 }.
\end{equation}

The effective vertices are known \cite{BFKL} \cite{FS} since long. The
projection on helicity states and the use of the complex notation lead
to an essential simplification. Introducing appropriate fields one can
write down the multi-Regge effective action \cite{KLS} involving these
vertices as interaction terms. It allows to reproduce in the easiest way
the leading contributions in the Regge limit of QCD. This effective
action can be derived from the original QCD action by separating the
momentum modes of the gluon and the quark fields according to the
multi-Regge kinematics and integrating out some "heavy" modes. There are
remarkable relations between interaction terms describing scattering and
production and between pure gluonic terms and those involving quarks.
The longitudinal and the transverse dimensions are clearly separated in
this effective action.

The case of two reggeized gluons is just the leading $\ln x$
approximation discussed above. In order to satisfy the conditions of
unitarity in the $s$-channel and in all sub-energy channels one has to
include the contributions  of arbitrary numbers of reggeons. This leads
us to the generalized leading logarithmic approximation:  The gluons in
all $s$-channel intermediate states should obey the multi-Regge
kinema\-tics Eq. (5) and the effective vertices and the trajectories
are taken in the leading logarithmic approximation Eq. (10), Eq.(11).
Beyond the leading logarithmic approximation we encounter other
contributions which arise e.g. if two s-channel gluons do not have a
large invariant mass. They have to be accounted for in later steps as
corrections to the trajectories and the effective vertices \cite{FL}.

The partial-wave equation describing the exchange of $r$ reggeons is a
straightforward generalization of the BFKL equation \cite{JB} \cite
{KwP} \cite{L89}. It is represented graphically in Fig. 6.

Both the
inhomogeneous term corresponding to $r$ non-interacting reggeons and the
interaction term involve the angular momentum part of the $r$-reggeon
pro\-pa\-ga\-tor, $[\omega - N \sum_{\ell} \alpha_G (\kappa_{\ell } )
]^{-1}$.
Both the trajectories Eq. (10) in this factor as well as the bare
reggeon
interaction kernel Eq. (12) involve infrared divergencies. In the
colour-singlet channel , which is the one of physical interest, these
divergencies cancel. We multiply the equation by the inverse of this
factor and include the terms involving $\alpha_G (\kappa )$ into the
kernel,
\begin{equation}
{\cal H} (\kappa_1, \kappa_2; \kappa_1^{\prime }, \kappa_2^{\prime}) =
\vert \kappa_1 \vert^{-2} \vert \kappa_2 \vert^{-2}
{\cal H}^{(0)} (\kappa_1, \kappa_2; \kappa_1^{\prime },
\kappa_2^{\prime}) -
\delta (\kappa_1 - \kappa_1^{\prime } )
\left (\alpha_G (\kappa_1 ) + \alpha_G (\kappa_2 ) \right ),
\end{equation}
which defines now a finite operator.

%Fig. 6

\begin{picture}(50000,21000)
\drawline\fermion[\E\REG](3000,5000)[5000]
\drawline\fermion[\N\REG](\particlebackx,\particlebacky)[10000]
\drawline\fermion[\W\REG](\particlebackx,\particlebacky)[5000]
\drawline\fermion[\S\REG](\particlebackx,\particlebacky)[10000]
\put(5000,10000){$ f $}
\drawline\fermion[\W\REG](3000,14000)[3000]
\drawline\fermion[\W\REG](3000,12000)[3000]

\drawline\fermion[\W\REG](3000,8000)[3000]
\drawline\fermion[\W\REG](3000,6000)[3000]

\drawline\fermion[\E\REG](8000,14000)[3000]
\drawline\fermion[\E\REG](8000,12000)[3000]

\drawline\fermion[\E\REG](8000,8000)[3000]
\drawline\fermion[\E\REG](8000,6000)[3000]

\put(13000,10000){$ = $}
\drawline\fermion[\E\REG](15000,14000)[6000]
\drawline\fermion[\E\REG](15000,12000)[6000]

\drawline\fermion[\E\REG](15000,8000)[6000]
\drawline\fermion[\E\REG](15000,6000)[6000]
\put(23000,10000){$ + $}
\drawline\fermion[\E\REG](32000,5000)[5000]
\drawline\fermion[\N\REG](\particlebackx,\particlebacky)[10000]
\drawline\fermion[\W\REG](\particlebackx,\particlebacky)[5000]
\drawline\fermion[\S\REG](\particlebackx,\particlebacky)[10000]
\put(34000,10000){$ f $}
\drawline\fermion[\W\REG](32000,14000)[5000]
\drawline\fermion[\W\REG](32000,12000)[5000]

\drawline\fermion[\W\REG](32000,8000)[5000]
\drawline\fermion[\W\REG](32000,6000)[5000]

\drawline\fermion[\E\REG](37000,14000)[3000]
\drawline\fermion[\E\REG](37000,12000)[3000]

\drawline\fermion[\E\REG](37000,8000)[3000]
\drawline\fermion[\E\REG](37000,6000)[3000]

\drawline\photon[\N\REG](28000,8000)[4]
\put(28000,8000){\circle*{300}}
\put(28000,12000){\circle*{300}}

\put(26000,14000){$i_r$}
\put(41000,14000){$\overline i_r$}

\put(26000,12000){$j$}
\put(41000,12000){$\overline j$}

\put(26000,8000){$i$}
\put(41000,8000){$\overline i$}

\put(26000,6000){$i_1$}
\put(41000,6000){$\overline i_1$}

\put(1500 ,13500){.}
\put(1500 ,13000){.}
\put(1500 , 12500){.}
\put(1500 ,10500){.}
\put(1500 ,10000){.}
\put(1500 , 9500){.}
\put(1500 ,7500){.}
\put(1500 ,7000){.}
\put(1500 ,6500){.}

\put(9500 ,13500){.}
\put(9500 ,13000){.}
\put(9500 , 12500){.}
\put(9500 ,10500){.}
\put(9500 ,10000){.}
\put(9500 , 9500){.}
\put(9500 ,7500){.}
\put(9500 ,7000){.}
\put(9500 ,6500){.}

\put(18000 ,13500){.}
\put(18000 ,13000){.}
\put(18000 , 12500){.}
\put(18000 ,10500){.}
\put(18000 ,10000){.}
\put(18000 , 9500){.}
\put(18000 ,7500){.}
\put(18000 ,7000){.}
\put(18000 ,6500){.}

\put(30500 ,13500){.}
\put(30500 ,13000){.}
\put(30500 , 12500){.}
\put(30500 ,10500){.}
\put(30500 ,10000){.}
\put(30500 , 9500){.}
\put(30500 ,7500){.}
\put(30500 ,7000){.}
\put(30500 ,6500){.}

\put(38500 ,13500){.}
\put(38500 ,13000){.}
\put(38500 , 12500){.}

\put(38500 ,10500){.}
\put(38500 ,10000){.}
\put(38500 , 9500){.}
\put(38500 ,7500){.}
\put(38500 ,7000){.}
\put(38500 ,6500){.}

%\put(40000,0){\sl Fig. 5}
\end{picture}

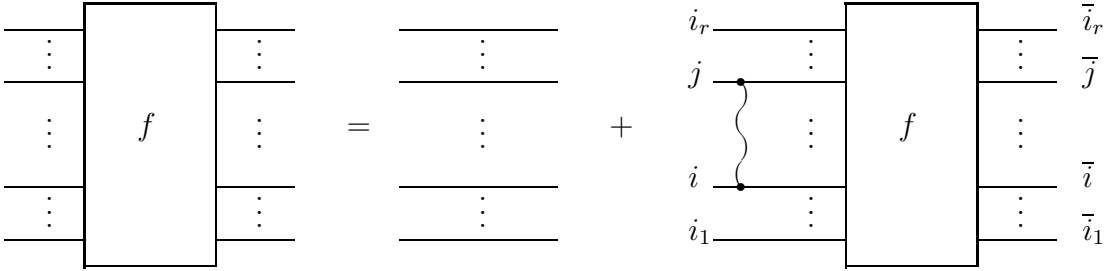
{\tenrm Fig. 6: \ \ Equation for the Green function of the $r$-reggeon
exchange.}
\vspace{1cm}

 We change to the impact parameter representation by Fourier
transformation with respect to the transverse momenta. The two-reggeon
interaction is now represented by an operator acting on the transformed
partial wave being now a function of the impact parameters $x_{\ell },
\ell = 1,...,r$.  The operator is obtained from Eq. (13) by applying the
substitutions
\begin{eqnarray}
&\kappa_1 \rightarrow \partial_1^*, \ \ \ \
\kappa_2^*  \rightarrow \partial_2, \cr
& \vert \kappa_1 -\kappa_1^{\prime } \vert^{-2} \rightarrow
- \ln \vert x_{12} \vert^2, \ \
 \alpha_G (\kappa_1 ) \rightarrow - \ln (\partial_1 \partial_1^* ) +
\psi (1).
\end{eqnarray}
$\partial_1 $ is the differentiation with respect to $x_1 $, $-\psi (1)
$ is the Euler - Mascheroni  number and $x_{12} = x_1 - x_2$.

It is
remarkable that the
resulting operator decomposes into a holomorphic and an antiholomorphic
part.
\begin{eqnarray}
&\hat {\cal H}_{GG} = H_G + H_G^*, \cr
& H_G = 2 \psi (1) - \partial_1^{-1} \ \ln x_{12} \ \partial_1
 - \partial_2^{-1} \ \ln x_{12} \ \partial_2
-\ln \partial_1 - \ln \partial_2.
\end{eqnarray}
In the case of two reggeized gluons we obtain
the homogeneous BFKL equation in the form
\begin{equation}
\omega f(\omega , x_1 ,x_2 ) = {g^2 N \over 8 \pi^2 } \hat {\cal H}_{GG}
f(\omega, x_1 ,x_2).
\end{equation}
 Because of the decomposition Eq. (15) this equation allows a
holomorphically factorized solution.

In the case $r \geq 4 $ the gauge group structure of the interaction
prevents in general such a factorization. However  in the large
$N$ approximation this factorization holds. Then the holomorphic part of
the $r$-reggeon equation Fig. 6 corresponds just to the Schr\"odinger
equation of a $r$-body system in one dimension with the interaction
restricted to the nearest neighbours and given by the unconventional
hamiltonian $- H_G $ in Eq. (15).  We shall see that it is really an
extraordinary hamiltonian.

We use operator relations expressing the derivative operator $\partial $
as a similarity transformation of $x^{-1}$,
\begin{equation}
\partial = \Gamma (x \partial + 1)^{-1} x^{-1} \Gamma (x \partial +1) =
  \Gamma (- x \partial ) x^{-1} \Gamma (- x \partial )^{-1},
\end{equation}
in order to obtain \cite{L93}
\begin{equation}
\ln \partial = \frac{1}{2} \left ( \psi (x \partial ) + \psi (1 - x
\partial ) \right )
+ (x \partial )^{-1} - \ln x .
\end{equation}
The commutation relations between $x$ and $\partial$ imply
\begin{equation}
\partial^{-1} \ \ln x \ \partial =
(x \partial )^{-1} \ \ln x \ (x \partial )  =
\ln x - (x \partial )^{-1}.
\end{equation}
Using Eqs. (18) and (19) we obtain
\begin{equation}
H_G = 2 \psi (1) - \frac{1}{2} \left (
\psi (x_{12} \partial_1 ) + \psi (- x_{12} \partial_2 ) +
\psi (1 - x_{12} \partial_1 ) + \psi (1 + x_{12} \partial_2 ) \right ).
\end{equation}
$\psi (z) $ is the digamma function,
\begin{equation}
\psi (1) - \psi (z) = \sum_{\ell = 0}^{\infty }
\left ( {1\over \ell + z }  - {1 \over \ell + 1 } \right ).
\end{equation}

We neglect the operator $x_{12} (\partial_1 + \partial_2 )$ in the
following transformation. This is justified in our kinematics, because
$i(\partial_1 + \partial_2 )$ is the momentum transfer operator and
$x_{12}$ is Fourier conjugated to $\kappa $. Then Eqs.(20) and (21)
imply \cite{L93}
\begin{eqnarray}
H_G = \chi_0 (x_{12}^2 \partial_1 \partial_2 ), \ \ \ \ \ \ \ \ \
 \chi_0 (z) = \sum_{\ell = 0}^{\infty} \left (
{2 \ell + 1 \over \ell (\ell + 1 ) + z } - {2 \over \ell + 1 }
\right ).
\end{eqnarray}
We have obtained $H_G$ as a function of the Casimir operator, $ x_{12}^2
\partial_1 \partial_2 $,  of holomorphic linear conformal
transformations acting on functions of $x_1$ and $x_2$. There is another
way to show that $H_G $ is conformally invariant. Therefore the result
Eq. (22) holds in general irrespective of the approximation done in the
last step.

The eigenvalues of the Casimir operator in the unitary representation
are $m (1-m) $, with $m = \frac{1}{2} + i \nu + \frac{n}{2} $. $n$ runs
over all integers and $\nu$ runs over the real axis. In this way we
obtain the eigenvalues of $H_G $. In particular we reproduce the
eigenvalues of the BFKL equation Eq. (15) and the leading behaviour Eq.
(2).

Including fermion exchange leads to an asymptotics down by powers of
$s$
compared to the exchange of gluons only. However in some channels the
quantum numbers exclude the latter contribution. For example, fermion
exchange contributions determine the small $x$ behaviour of flavour
non-singlet structure functions. The interaction between two reggeized
fermions and between a reggeized fermion and a reggeized gluon are
described in analogy to the gluon-gluon case dicussed here.There is a
peculiarity in the two-fermion case because of double-logarithmic
contributions \cite{RK941}. The properties of holomorphic factorization
and conformal symmetry hold for all two-reggeon interactions in QCD
\cite{RK942}.

 The holomorphic linear conformal transformations of the functions of
$x_1 $ and $x_2 $ are generated by $M_{12}^{a} = M_1^{a} + M_2^{a} $,
with $M_1^{+} = x_1^2 \partial_1, M_1^{-} = \partial_1, M_1^{(0)} = x_1
\partial_1 $. The commutation relations are the ones of the angular
momenta modified by a sign. We introduce the $2\times 2 $ matrix
$L_{\ell } (\theta )$, one for each reggeon, $\ell = 1,..., r$,
\begin{eqnarray}
L_{\ell } (\theta ) =
\left ( \matrix {\theta + M_{\ell }^{(0)}   & M_{\ell }^{-} \cr
- M_{\ell }^{+} & \theta - M_{\ell }^{(0)} \cr }
\right ).
\end{eqnarray}
The commutation relations of the generators $M^{a}_{\ell}$ can be
written in the $R$-matrix form \cite{L94},
\begin{equation}
\left ( L_{\ell} (\theta_1 ) \otimes L_{\ell } (\theta_2 ) \right )
R(\theta_1 - \theta_2 ) =
R(\theta_1 - \theta_2 ) \left (
 L_{\ell} (\theta_2 ) \otimes L_{\ell } (\theta_1 ) \right ).
\end{equation}
Here $\otimes $ denotes the tensor product. $R(\theta) $ is the
$4\times 4 $ matrix the non-vanishing entries of which are
$R_{1 1, 1 1} (\theta ) = R_{2 2, 2 2 } (\theta ) = \theta +1 ,
R_{1 2, 1 2} (\theta ) = R_{2 1, 2 1 } (\theta ) = 1 ,
R_{1 2, 2 1} (\theta ) = R_{2 1, 1 2 } (\theta ) = \theta $.
It obeys the Yang-Baxter relation \cite{YB}.
Therefore the operators   tr$ \prod_{\ell =1}^{r} L_{\ell } (\theta )
$
with $\theta = \theta_1 $ and $\theta = \theta_2 $ commute and generate
by decomposition in $\theta $ \ \ \ $(r-1) $ mutually commuting
operators.

We see that the problem of the $r$-reggeon exchange in the large $N$
limit leads to a periodic lattice spin model \cite{L94}, a
generalization of the Heisenberg XXX model. The spin $\frac{1}{2} $ at
each site in the Heisenberg model is here replaced by a representation
of the group SL(2,R). This lattice model is completely integrable
\cite{L94} \cite{FK}.

There is another $R$-matrix being an operator acting on functions of
$x_1 $ and $x_2 $ and obeying
\begin{equation}
\left ( L_1 (\theta_1 ) \cdot L_2 (\theta_2 ) \right )
{\bf R}(\theta_2 - \theta_1 ) =
{\bf R}(\theta_2 - \theta_1 ) \left (
 L_2 (\theta_2 ) \cdot L_1 (\theta_1 ) \right ).
\end{equation}
The dot denotes usual matrix multiplication. ${\bf R}(\theta) $
generates another set of commuting operators. They commute also with the
ones of the first set \cite{FK} \cite{YB}. The first non-trivial term in
the $\theta $-decomposition contains just the hamiltonian $-H_G$.
\begin{equation}
{\bf R}(\theta ) = P_{12} \left ( 1 - \theta H_G + {\cal O} (\theta^2)
\right ).
\end{equation}
With these results the problem of the $r$-reggeon exchange
asymptotics is
in principle solved in the large $N$ approximation. It remains to work
out the Improvements for $N=3$ and the phenomenological implication
concerning the saturation of the increase of the structure functions at
very small $x$.

\newpage

{\Large \bf Acknowledgements}

\vspace{1cm}

I thank Prof. B.F.L. Ward for inviting me to this interesting conference
at a pleasent place. I am grateful to the DESY Directorate, in
particular to Prof. P. S\"oding, for providing me generous support.
My understanding of the subject has been essentially improved
in the collaboration with L.N. Lipatov and L. Szymanowski. This
collaboration is supported in part by the Volkswagen-Stiftung.

\end{document}